# Lateral quantum confinement effect on monolayer high-T$_c$ superconductors

Guanyang He,[1] Yu Li,[1] Yuxuan Lei,[1,4] Andreas Kreisel,[5] Brian M. Andersen,[5] and Jian Wang[1,2,3,4*]

[1]*International Center for Quantum Materials, School of Physics, Peking University, Beijing 100871, China*
[2]*Collaborative Innovation Center of Quantum Matter, Beijing 100871, China*
[3]*Hefei National Laboratory, Hefei 230088, China*
[4]*Beijing Academy of Quantum Information Sciences, Beijing 100193, China*
[5]*Niels Bohr Institute, University of Copenhagen, 2100 Copenhagen, Denmark*

(Dated: December 30th, 2023)

Despite decades of research in spatially confined superconducting systems to understand the modification of superconductivity from reduced length scales, the investigation of the quantum confinement effect on high-temperature superconductors remains an outstanding challenge. Here, we report scanning tunneling spectroscopy measurements on laterally confined FeSe monolayers on SrTiO$_3$ substrates, which are formed by epitaxially growing FeSe films with a coverage less than one unit cell. Comparing to the uniform regions of FeSe monolayers, the peninsula regions at the monolayer boundary exhibit reduced Fermi energy and undiminished superconductivity, leading to a putative crossover from a Bardeen-Cooper-Schrieffer state to a Bose-Einstein condensate state. In isolated FeSe monolayer islands, superconductivity is shown to exist in samples of smaller volume in contrast to conventional superconductors, while the validity of Anderson's criterion remains fulfilled. Our work reveals lateral quantum confinement effects in unconventional superconductors, to enrich the understanding of high-temperature superconductivity in low-dimensional systems.

**Introduction.** When the size of a superconductor goes below or becomes comparable to certain characteristic length scales, such as the superconducting (SC) coherence length ξ, quantum confinement can modify SC properties and generate novel phenomena. Decades of active research has been fueled by the quantum confinement effect on superconductivity, and especially enriched by the study of two-dimensional (2D) superconductors with thicknesses less than ξ. Nowadays, 2D superconductors have become an important platform to study quantum phase transitions and other important quantum behavior [1-9]. As the area of 2D superconductors decreases, lateral quantum confinement can also affect the local electronic structure and superconductivity.

Despite extensive research in quantum confinement effects, the lateral quantum confinement in 2D high critical temperature (T$_C$) superconductors remains unexplored. Considering the small ξ of high-T$_C$ superconductors [10], it is particularly challenging to approach these characteristic lengths laterally. In this work, to investigate such quantum confinement by scanning tunneling microscopy/spectroscopy (STM/S), we grew monolayer FeSe on SrTiO$_3$(001) (STO) substrate with a coverage less than one unit cell, to generate laterally confined monolayer FeSe films and islands. The 0.55 nm thickness of FeSe monolayers is below its coherence length ξ ~1.2 nm. Two kinds of laterally confined FeSe monolayer are studied. The first is FeSe peninsulas near the film boundaries with confined widths around ten nanometers, where the SC gap remains intact, but the Fermi energy ($E_F$) is reduced significantly. Accordingly, the SC pairing strength $\Delta/E_F$ varies from 0.28 for uniform FeSe monolayers to 0.73 for peninsulas, indicating a putative crossover from the Bardeen-Cooper-Schrieffer (BCS) regime to the Bose Einstein condensation (BEC) regime. The second is isolated monolayer FeSe islands with areas around tens of square nanometer, where a coexistence of Coulomb gap and SC gap is observed. Upon decreasing the size of the FeSe islands, the SC gap is reduced and eventually disappears. Our findings turn out consistent with the Anderson criterion, i.e. that for small enough superconductors, the electronic energy level spacing destabilizes SC order [11].

**Methods.** Our experiments are conducted in a molecular beam epitaxy (MBE) - STM combined system (Scienta Omicron, Inc.) with an ultrahigh vacuum of $1\times10^{-10}$ mBar, where the FeSe monlayer is epitaxially grown on Nb-doped $SrTiO_3$ (001) (wt 0.7%) substrates (STO). STO is thermally annealed in vacuum at 1050 °C for 40 minutes to obtain an atomically flat $TiO_2$-terminated surface. The FeSe monolayer is grown by co-evaporating high-purity Fe (99.994%) and Se (99.999%), with the substrate held at 400 °C. Then, the as-grown FeSe film is annealed at 450 °C for 2 hours.

The STS data is measured *in situ* in the STM at 4.2 K by a polycrystalline Pt/Ir tip and standard lock-in technique. The modulation voltage on the tip is 1 mV at 1.769 kHz. The set-up of STS measurements is $V$ = 40 mV, $I$ = 2.5 nA for tunneling spectra, and $V$ = 1 V, $I$ = 0.2 nA for topographic images unless specified otherwise.

**Results and Discussions.** The STM topographic image of our monolayer FeSe grown on the STO terraces is shown in Fig. 1a, and the inset provides the top Se atom arrangement of it. In Fig. 1b, a symmetrized d$I$/d$V$ tunneling spectrum is obtained from STS measurements in the central uniform area of monolayer FeSe (see Supplementary for spectra symmetrization), exhibiting prominent coherence peaks and U-shaped SC gaps at 4.2 K. The two-band Dynes model with anisotropic SC gap functions fits the d$I$/d$V$ spectrum well to give two gap values $\Delta_1$ = 10.3 meV and $\Delta_2$ = 18.1 meV [12].

Quasiparticle interference (QPI) in the differential conductance mapping $g(r, E) = dI/dV(r, eV)$ provides a powerful tool to analyze electronic states [13-15]. First, $g(r, E)$ are obtained on FeSe monolayers. After Fourier transform (FT), the modulus of FT-QPI $|g(q, E)|$ as the scattering intensity exhibits three types of ring-like scattering structures, denoted by $q_1$, $q_2$ and $q_3$ rings in Fig. 1c; $q$ is the momentum transfer of the scattering momenta $k$ on the Fermi pockets ($q = k_1 - k_2$). Due to the tunneling-matrix-element effect and the orbital structure of the Fermi pockets, the $q_1$ ring exhibits the highest scattering intensity and most complete ring shape [12,16], therefore is used to extract the band dispersion. Since the elastic scattering from the opposite side of the pocket ($-k \leftrightarrow +k$) normally results in the highest intensity at $q = 2k$ [14], the band dispersion is approximately $E(k) = E(q/2)$. The energy evolution of the $q_1$ ring is resolved from maximum-intensity points in a $|q|$-$E$ plane, corresponding to the dark points on both sides of the $|q|$ = 0 position in Fig. 1d. Here, as the $q_1$ ring corresponds to the scattering within an electron pocket, the ring size shrinks as the energy decreases. The red dashed line in Fig. 1d is a parabolic fitting for the band dispersion (dark points), and $E_F$ is extracted from the distance between the bottom of the parabola and the Fermi level [15,17]. To present a sharper image, the contributions to the scattering intensity by other electronic pockets inside the parabola are manually subtracted (see Supplementary for QPI analysis). With more QPI measurements in other uniform areas (Fig. S6 and Table S1), an average of $E_F$ = 65.3 meV is obtained, close to results in the literature [17].

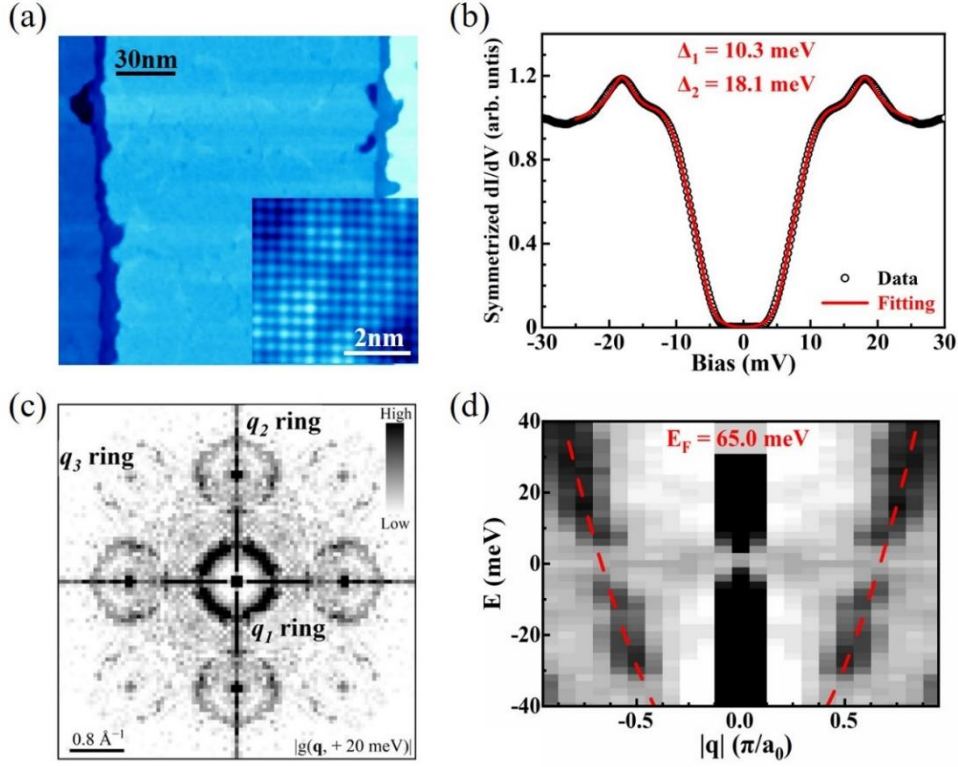

**Fig. 1 STM topography, tunneling spectrum and band dispersion of the uniform area of FeSe monolayers.** (a) Large-scale STM topographic image of the FeSe monolayer grown on STO terraces, where darker colors mean lower heights. The inset is an atomically resolved image showing the topmost Se lattice of FeSe. (b) A typical symmetrized $dI/dV$ spectrum measured at 4.2 K in the uniform area of a FeSe monolayer. The red curve is the theoretical Dynes fit. (c) Typical FT-QPI pattern of $|g(\boldsymbol{q}, E = 20 \text{ meV})|$ after symmetrization, showing three types of ring-like structures denoted by $\boldsymbol{q_1}$, $\boldsymbol{q_2}$ and $\boldsymbol{q_3}$ rings. (d) Intensity plot in the $|\boldsymbol{q}|$-$E$ plane, where $|\boldsymbol{q}| = 0$ corresponds to the center of (c). The intensity at each energy is plotted versus the radial coordinate near the $\boldsymbol{q_1}$ ring position after azimuthal averaging, showing the energy evolution of the $\boldsymbol{q_1}$ ring. $a_0 = 0.38$ nm is the lattice constant of FeSe. The red dashed line is a parabolic fitting for the maximum-intensity points at each energy to find $E_F$.

Figure 2a shows a peninsula structure at the boundary of a FeSe monolayer, spatially confined in the width as indicated by the double arrow; the dark area is the STO substrate. In the dashed box region, QPI measurements are performed again to extract the band dispersion (Fig. 2b), and the dI/dV tunneling spectrum is measured (Fig. 2c). The resolution of $|q|$ from a QPI area of L×L is $2\pi/L$. Considering a minimum value of L = 7.8 nm in Fig. 2a, the $|q|$ resolution is around 0.1 ($\pi/a_0$), which appears sufficient to reveal the band dispersion in Fig. 2b. By the same method to determine $E_F$ in the previous paragraph, $E_F$ in this peninsula appears much smaller than that in the uniform area by nearly 40 meV (from 65.3 to 26.8 meV). The energy band becomes flatter in the peninsulas with a decreased Fermi velocity $v_F = \frac{1}{\hbar}\frac{\partial E}{\partial k} \approx 2.8 \times 10^4$ m/s (calculated from Fig. 2b), which is smaller than $v_F \approx 7.1 \times 10^4$ m/s in the uniform area (calculated from Fig. 1d). Meanwhile, the Dynes fitting of SC gaps in Fig. 2c gives $\Delta_1 = 9.3$ meV and $\Delta_2 = 19.5$ meV, which barely change from those in the uniform area ($\Delta_1 = 10.2$ meV, $\Delta_2 = 18.1$ meV), considering the possible gap value fluctuation in monolayer FeSe/STO of around 1.5 meV [18]. Figure 2d-f show the results of another peninsula, which corroborates the reduction of $E_F$ and the undiminished SC gaps. Similar results from many more peninsulas are exhibited in the supplementary Fig. S8 and S9, and a summarization of all $\Delta$ and $E_F$ values is given in Fig. 2g and Table S2. For each point in Fig. 2g, the horizontal error bar represents the standard deviation of the parabolic fitting for $E_F$, and the vertical error bar represents the standard deviation of the Dynes fitting for SC gaps. Moreover, the temperature dependence of the tunneling spectra is measured from 4.2 K to 42 K, and the BCS fitting for $\Delta_{1,2}(T)$ gives $T_c \approx 49$ K (see Fig. S7 in Supplementary). Thus, $T_c$ is nearly unchanged from the uniform area to the peninsulas.

Generally speaking, superconducting $T_c$ and $\Delta$ are supposed to decrease sharply in the presence of quantum confinement because of the Anderson criterion [19]. On the other hand, for monolayer FeSe/STO, it is believed that an increased carrier density in FeSe (due to electron doping from STO) can boost superconductivity [20-22]. In our monolayer FeSe peninsulas, the effective mass of the electrons $m^* = \frac{\hbar^2|k_F|^2}{2E_F}$ is larger than that in the uniform area because of the reduced $E_F$ (see Fig. S10b in Supplementary for $m^*$). Since the carrier density of states is proportional to $m^*$ for two-dimensional systems, such enlarged $m^*$ because of quantum confinement could suggest the boosted superconductivity as well. Superconductivity in the unconventional superconductors Al, In, and Sn nano-particles is enhanced with the decrease of sample sizes [23,24], which is ascribed to phonon softening caused by the small sample size, specifically by structural changes in lattice or contributions from surface phonons [11,25,26]. In our case, the factors related to quantum confinement mentioned above may counteract with each other to a certain extent, so that $\Delta$ and $T_c$ remain nearly unchanged in the peninsulas as discussed in the previous paragraph. Moreover, a stronger electron correlation is expected due to the decreased $v_F$ (larger $m^*$) in the peninsulas, resulting in the BCS-BEC crossover that will be discussed in the next paragraph.

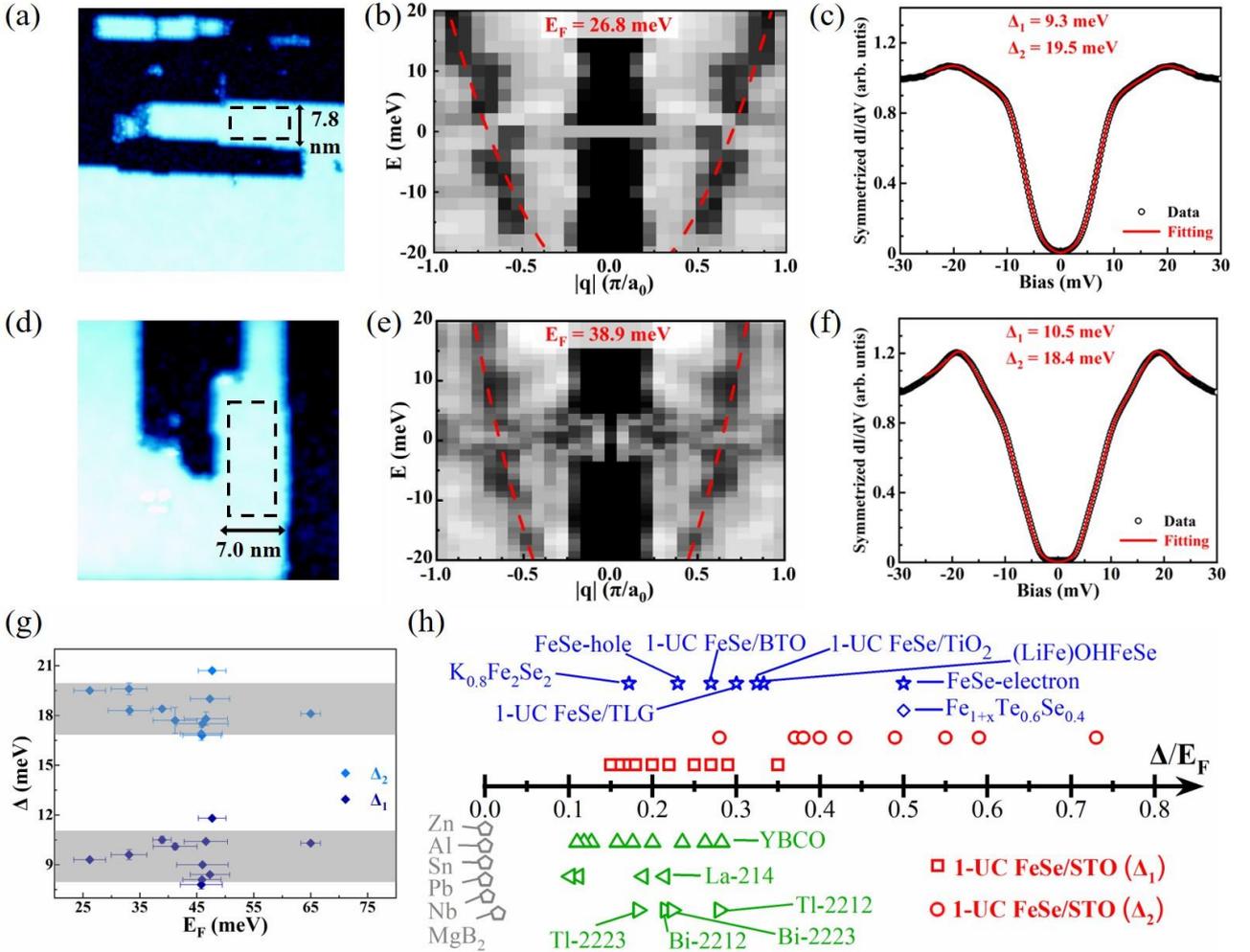

**Fig. 2 STM topographies, band dispersions and tunneling spectra in the peninsulas of FeSe monolayer.** (a) STM topographic image of a peninsula at the monolayer boundary. The dashed box indicates the area of QPI analysis, and the double arrow indicates the peninsula width. (b) Band dispersion extracted from the QPI patterns in the boxed region of (a) by a parabolic fitting. (c) Symmetrized tunneling spectrum measured at the center of the dashed box in (a) with a Dynes fitting. (d-f) Similar results to (a-c) for another peninsula. (g) The summarized results of $\Delta$ and $E_F$ from different peninsulas (peninsula widths from 6 to 27 nm). (h) Multiple SC systems in the BCS limit and BCS-BEC crossover regime. Grey pentagons represent the results from conventional superconductors. Green symbols represent cuprates. Blue symbols represent iron-based superconductors. Red symbols represent our results in (g).

The microscopic theory of superconductivity is based on the formation of fermion pairs. With a weak attractive interaction between fermions, they pair in momentum space with a large pair size $\xi$, classified as the BCS limit. With a strong attractive interaction, fermions will be bound in pairs tightly in real space; such preformed pairs undergo BEC to the superfluid state afterwards, classified as the BEC limit. These two limits are connected through an intermediate unitary regime called BEC-BCS crossover [27,28]. This crossover is important for understanding strongly correlated electronic systems, attracting attention in cold atoms and condensed matter fields. In cold atom systems, such crossover is realized by tuning the interaction strength between atoms. The interaction strength is specified by the scattering length $a_s$, which diverges at the BEC-BCS crossover. Also, the ratio of $\Delta/E_F$ is around 0.5 as a hallmark of this crossover [29]. In condensed matter systems, $\Delta/E_F$ is studied in various superconductors as we summarize in Fig. 2h. For conventional superconductors Al, Sn, and Nb, $\Delta/E_F$ is nearly zero [30], and for MgB$_2$ the value of $\Delta/E_F$ is 0.016 [31], all suggesting BCS limit. On the other hand, BCS-BEC crossover is discussed in high-$T_c$ cuprates due to the quasi-2D nature and strong pairing [32]. $\Delta/E_F$ values of cuprates are higher than those of conventional superconductors, but mostly lower than 0.3. Specifically, the upward triangles in Fig. 2h refer to YBa$_2$Cu$_3$O$_x$, leftward triangles to La$_{2-x}$Sr$_x$CuO$_4$, and rightward triangles to Tl$_2$Ba$_2$Ca$_2$Cu$_3$O$_{10}$, Bi$_2$Sr$_2$CaCu$_2$O$_8$, Bi$_2$Sr$_2$Ca$_2$Cu$_3$O$_{10}$, Tl$_2$Ba$_2$CaCu$_2$O$_8$ [33-35]. In iron-based superconductors, the effects of chemical doping have been explored, and $\Delta/E_F$ of bulk Fe$_{1+x}$Te$_{0.6}$Se$_{0.4}$ can be raised from 0.16 to 0.50 by reducing x [36]. Different substrates have also been used to tune the band edge of FeSe monolayers; a FeSe monolayer grown on trilayer graphene (TLG) features $\Delta/E_F$ = 0.3 [37]. Bulk FeSe exhibits uniquely tiny hole ($E_F$ ~ 10 meV) and electron pockets ($E_F$ ~ 3 meV) at the Fermi surface [13], and $\Delta$ = 2.3 or 1.5 meV for the hole or electron band, respectively [38]. Thus $\Delta/E_F$ ~ 0.23 or 0.5 for the hole or electron band of bulk FeSe. In our work, monolayer FeSe/STO has much larger $T_c$ and yet a larger $E_F$, yielding $\Delta_2/E_F$ = 0.28 in the uniform area of our sample; this ratio is dramatically raised to 0.73 due to the lateral quantum confinement of the peninsula area. The BCS-BEC crossover in the peninsula could be pertinent to the enhanced electron correlation discussed in the previous paragraph, as strongly correlated electrons tend to condense into BEC-like pairs [39]. Therefore, monolayer FeSe peninsulas constitute a promising platform to study the BCS-BEC crossover modulation in condensed matter systems. The BEC regime might be considered to associate with pseudo-gap behavior from preformed Cooper pairs [40], while our peninsulas are in the crossover regime and no evidence of such pseudo-gaps is found as the temperature is raised from 4.2 to 42 K (Fig. S7).

Next, we focus on isolated islands of monolayer FeSe as shown in Fig. 3a, which are more spatially confined comparing to the peninsulas. When the size of a nano-island is small enough, electrons inside the island create a strong Coulomb repulsion preventing the tunneling of an external electron, known as Coulomb blockade. When the applied voltage overcomes the electrostatic energy, external electrons can tunnel into the island one at a time, leading to equidistant conductance peaks in the differential conductance spectrum known as the Coulomb staircase [41]. Figure 3b shows the tunneling spectrum measured on the FeSe island in Fig. 3a. Except for the gap at the Fermi surface (i.e., zero bias) which will be discussed later, pronounced differential conductance oscillations with nearly equidistant peaks is observed. The peak spacing is labeled by "U" in Fig. 3b. This observed phenomenon is consistent with the scenario of Coulomb staircase peaks, where the voltage interval U is related to the addition energy of an electron to the island. $U = e/C_1$ depends on the capacitance $C_1$ between the island and STM tip [42,43], and the observation of U = 14.0 mV (Fig. 3b) gives $C_1$ = 11.4 aF. The black dots in Fig. 3c summarize the values of U in FeSe islands of different sizes (see more results in Fig. S11). For each dot, the error bar represents the standard deviation of U from multiple Coulomb staircase peaks measured on one island. These dots follow a linear relation (red dashed line) between U and 1/S, confirming the relationship $U = e/C_1 \propto 1/S$, S as the area of monolayer FeSe islands.

The inset of Fig. 3b shows an enlarged view at zero bias, where a gap appears bigger than U and asymmetric about zero bias. The asymmetry about zero bias can be attributed to the residual charge on the island. According to orthodox Coulomb blockade theory [42,43], the Coulomb blockade gap ($e/C_2$) due to the island-substrate capacitance $C_2$ is expected to show up at zero bias together with the gap of $U = e/C_1$, which is not our case. Our observation of a single gap at zero bias (Fig. 3b) may suggest that $C_2$ is much larger than $C_1$ and thus the Coulomb blockade gap ($e/C_2$) is negligible, considering the tiny area of STM tip. Interestingly, the zero-bias gap does not exhibit a monotonous relation with S (the area of monolayer FeSe islands) in Fig. 3d. Since either the gap of $e/C_1$ or $e/C_2$ is supposed to be inversely proportional to S [43], the Coulomb interaction alone cannot explain the observed zero-bias gap. Given that the SC gap shrinks for a decreasing sample volume (proportional to S in our case) near the Anderson limit [19], the zero-bias gap could be a mixture of U and the SC gap, as the intrinsic energy gap (SC gap) of an isolated system can add to the Coulomb gap in the dI/dV spectrum [44,45]. We note within a typical capacitively-coupled superconducting island scenario, there is an expected even-odd effect arising from paired vs. unpaired electrons [19]. Nevertheless, we have only observed weak fluctuations in the addition energy of electrons between even and odd conductance peaks (see Fig. S12 in Supplementary). The absence of a clear even-odd effect for the FeSe islands studied in this work may be related to the reduction of superconductivity in the biased case due to the low critical superfluid density of the islands.

At zero bias voltage (low tunneling current), similar to the addition of U and the SC gap in Sn nanoparticles and Pb islands [46,47], we can calculate the underlying SC gap Δ with the assumption that the enlarged energy gap equals U + 2Δ [47]. The calculated Δ shown as black dots in Fig. 3e reduces with the decrease of FeSe island volume, where the error bar indicates the standard deviation of U. The red dashed line is a guide to the eye. Such behavior can be expected since small superconductors experience large thermal fluctuations in the order parameter which quenches superconductivity [11,48-50]. Moreover, the Anderson criterion suggests that SC state will completely disappear, when the mean electronic energy level spacing (Kubo gap) δ near $E_F$ exceeds the SC gap [51]. In a system of volume $V$, δ is around $\frac{2\pi^2\hbar^2}{m^*k_F V}$, where $k_F$ is the Fermi wave vector and $m^*$ the effective electron mass [52]. For monolayer FeSe with or without lateral confinement, $k_F$ is both around 2.7 nm$^{-1}$ (Fig. S10a). $m^*$ is enlarged in the presence of lateral confinement for peninsulas (Fig. S10b), and could be even larger in the case of isolated islands. Thus, by adopting the largest $m^* = 10.9\, m_e$ in Fig. S10b, δ vs $V$ is plotted in Fig. 3e as the blue dashed line. The Anderson limit is estimated by the intersection between two dashed lines in red and blue colors where $δ = Δ$, giving a critical volume of 29 $nm^3$. In contrast to conventional superconductors with Anderson limits around 64 ~ 216 $nm^3$ [11], superconductivity persists in monolayer FeSe islands down to a smaller volume. The critical volume of 29 $nm^3$ is also consistent with the fact that we do not detect any SC gap in islands smaller than 29 $nm^3$ (Fig. S11d).

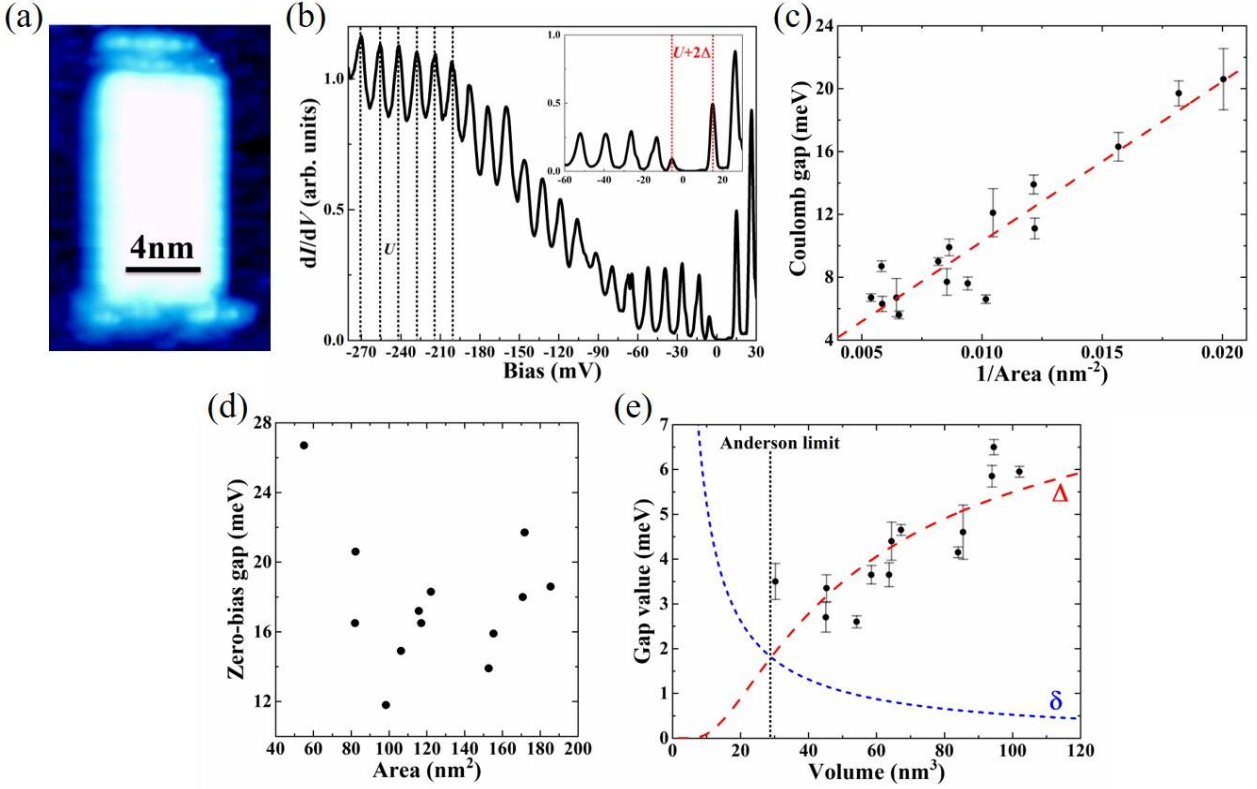

**Fig. 3 Coexistence of Coulomb gap and SC gap in isolated monolayer FeSe islands.** (a) Topographic image of a rectangular monolayer FeSe island on STO substrates. (b) Tunneling spectrum measured on the island in (a). The periodic gap $U$ away from zero bias is the Coulomb gap. The enlarged gap at zero bias consists of the Coulomb gap $U$ and SC gap. (c) Coulomb gaps in relation to the reciprocal of the island area, with the dashed line showing a linear fitting. (d) Dependence of the enlarged gap at zero bias on the island area. (e) SC gap value vs the volume of FeSe islands. The red dashed line is a guide to the eye for SC gap $\Delta$, and the blue dashed line shows the Kubo gap $\delta$. Anderson limit is estimated by the intersection between two dashed lines.

In conclusion, our work reveals how lateral quantum confinement affects the unconventional superconductivity of monolayer FeSe/STO in peninsula structures and isolated islands. In our experiments, lateral quantum confinement brings a clear reduction in $E_F$, but nearly undiminished superconductivity in FeSe peninsulas. A potential BCS-BEC crossover is thus detected, indicating a scenario that fermions form pairs before condensation in such low-dimensional high-$T_c$ superconductors. The pairing strength $\Delta/E_F$ is tuned by lateral quantum confinement from 0.28 to 0.73, suggesting a new method to reach the BCS-BEC crossover regime. For the first time, our work reveals that QPI analysis can be used to study the quantum confinement effect on low-dimensional superconductors. As the lateral confinement becomes more severe in isolated monolayer FeSe islands, superconductivity is found to be suppressed and coexists with Coulomb blockade effect. The evolution of the SC gap with the volume of monolayer FeSe islands is investigated, where superconductivity eventually disappears near the Anderson limit of $29\ nm^3$. This limit is much smaller than that in conventional superconductors, demonstrating that the Anderson limit is a valid criterion also in unconventional superconductors.

This work was supported by the National Key Research and Development Program of China (Grant No. 2018YFA0305604), the National Natural Science Foundation of China (Grant No. 11888101), the Innovation Program for Quantum Science and Technology (2021ZD0302403).

# Supplementary Information for

# Lateral quantum confinement effect on monolayer high-T$_C$ superconductors

Guanyang He,[1] Yu Li,[1] Yuxuan Lei,[1,4] Andreas Kreisel,[5] Brian M. Andersen,[5] and Jian Wang[1,2,3,4*]

[1]*International Center for Quantum Materials, School of Physics, Peking University, Beijing 100871, China*
[2]*Collaborative Innovation Center of Quantum Matter, Beijing 100871, China*
[3]*Hefei National Laboratory, Hefei 230088, China*
[4]*Beijing Academy of Quantum Information Sciences, Beijing 100193, China*
[5]*Niels Bohr Institute, University of Copenhagen, 2100 Copenhagen, Denmark*

**CONTENTS**

**STS data processing**

Originating from the band edge effect [53], the tunneling spectrum can be asymmetric with a higher density of states at positive bias voltages than at negative bias voltages. To eliminate such irrelevant effect, our raw STS spectrum (Fig. S1a) is normalized by dividing it by its asymmetric background (Fig. S1b), and the background curve is acquired by a cubic fitting to the raw spectrum in the voltage range away from the Fermi level (25 ~ 40 mV). This data processing method is previously used in the literature [18,54,55]. After the normalization, the particle-hole symmetrization (Fig. S1c) is performed in the next step by averaging all dI/dV values at -$V$ and +$V$ for each |$V$|, to give the symmetrized dI/dV spectra shown in the main text.

**Dynes Fitting and BCS Fitting for STS spectra**

The symmetrized spectrum is fitted by the summation of two Dynes functions ($\frac{dI_1}{dV}$ and $\frac{dI_2}{dV}$) with different weights ($w$): [56,57]

$$G = \left[w\frac{dI_1}{dV} + (1-w)\frac{dI_2}{dV}\right]$$

where the Dynes function with anisotropic s±-wave gap structure is given by [58,59]

$$\frac{dI_{1,2}}{dV} = N(E_F)\frac{1}{2\pi}\frac{1}{k_BT}\int_{-\infty}^{+\infty} dE \int_0^{2\pi} d\theta \, Re\left[\frac{|E - i\Gamma_{1,2}|}{\sqrt{(E - i\Gamma_{1,2})^2 - \Delta_{1,2}(\theta)^2}}\right] \cosh^{-2}\frac{E + eV}{2k_BT}$$

$$\Delta_{1,2}(\theta) = \Delta_{1,2}^{max}[1 - p_{1,2}(1 - \cos 4\theta)]$$

In the fitting parameters, $N(E_F)$ represents the density of states at $E_F$; $\Gamma$ is the scattering rate stemming from finite-lifetime effects of the quasiparticles at the gap edge [5,6]. Our STS spectra give $\Gamma \approx 2$ meV at 4.2 K, which is reasonable compared to our large superconducting gaps up to 18 meV. $p$ is the degree of anisotropy, which should be below 0.5 for nodeless superconducting gaps. Our STS spectra give $p \approx 0.3$ at 4.2 K, indicating a noticeable gap anisotropy. $\Delta_{1,2}^{max}$ represents the superconducting gap size, corresponding to the $\Delta_1$ and $\Delta_2$ discussed in the main text. For temperature-dependent tunneling spectra, $\Delta_1$ and $\Delta_2$ in the main text extracted by the Dynes fittings are further fitted by the BCS gap function to give $T_c$:

$$\Delta(T) = \Delta_0 \tanh\left(\frac{\pi}{2}\sqrt{\frac{T_c}{T} - 1}\right)$$

**Quasiparticle Interference Measurements and Band Extraction.**

After the differential conductance mapping $g(r, E)$ at different energies ($E$) is measured on the uniform area of monolayer FeSe (center of Fig. 1a), Fig. S2 exemplifies the data processing to obtain a sharp FT-QPI pattern. We use Lawler-Fujita Algorithm to correct the distortion of $g(r, E)$ [60]. Such correction is firstly performed on the topographic image simultaneously measured with $g(r, E)$, where the best correction parameters are found that eliminate all distortions in the known lattice structure of FeSe. Fig. S2a and S2b show the topographic image and the corresponding FT pattern respectively, which exhibit the non-orthogonality of the *x-y* axes and noticeable noise around the Bragg spots. After Lawler-Fujita correction, the topography (Fig. S2c) and the corresponding FT pattern (Fig. S2d) become a strictly tetragonal lattice. Then, the correction parameters for the topography are used in the following Lawler-Fujita correction for $g(r, E)$. The initial $g(q, E)$ (Fig. S2e) from the FT of $g(r, E)$ is mirror symmetrized and four-fold rotational ($C_4$) symmetrized (Fig. S2f), in order to increase the signal-to-noise ratio. These symmetrization methods are commonly used for the FT-QPI data processing in 1-UC FeSe [61,62]. Next, we use the $g(q, 0\,meV)$ to normalize $g(q, E \neq 0\,meV)$ in Fig. S2f, to suppress the intensity of tiny scattering vectors around the center (Fig. S2g). Note that this normalization is similar to a Gaussian core subtraction (Fig. S2h), while the latter requires manually inputting the parameters of Gaussian core.

After the above process to obtain Fig. S2g, we take line-cuts of $|g(q, E)|$ along $\theta = 22.5°$ and plot the intensity in a $|q|$-$E$ plane (Fig. S3b), which clearly reveals the electron band structure. We also perform the azimuthal averaging (19 line-cuts evenly spaced between 0°~45°) to further increase the signal-to-noise ratio (Fig. S3c). As for line-cuts from the initial $|g(q, E)|$ without symmetrization or normalization (Fig. S3a), the trend of the band dispersion can be roughly resolved but with poor signal-to-noise ratio. The same data process to obtain Fig. S3c is exhibited in the results of the peninsula region on monolayer FeSe (Fig. S4 and S5), and used in all results of the main text.

Finally, the band dispersion is obtained from the maximum-intensity points in a $|\boldsymbol{q}|$-$E$ plane, with a parabolic fitting:

$$E = -E_F + \frac{\hbar^2}{2m^*} \cdot |\boldsymbol{k}|^2, |\boldsymbol{k}| = |\boldsymbol{q}|/2$$

$E_F$ is the Fermi energy and m$^*$ is the effective mass. The band dispersion is fitted with both m$^*$ and $E_F$ as free parameters, to extract the $E_F$ values shown in Fig. 2g, Table. S1, and Table. S2.

In addition, we have also tried to fix m$^*$ = 4m$_e$ (the m$^*$ value in Fig. S10b for the uniform area of FeSe monolayer) with only $E_F$ as free fitting parameters. This m$^*$ is close to the value we deduce (~ 4.26 m$_e$) from the reference [61]. Fittings with fixed m$^*$/free m$^*$ parameter are both shown in Fig. S8, S9 by blue/red parabolic curves, where the maximum-intensity points in the $|\boldsymbol{q}|$-$E$ plane are denoted by white circles. The goodness of each fitting is determined by the $\chi^2$ value (Pearson's chi-square test) summarized in Table. S3, where the fittings with free m$^*$ parameters appear to be better than those with fixed m$^*$, suggesting that m$^*$ differs between the FeSe peninsula and the uniform area. Pearson's chi-square is calculated by $\chi^2 = \sum_i \frac{(E_i - E_{i0})^2}{E_{i0} + 50}$, where $E_i$ is the energy of the maximum-intensity point (y-axis coordinate of each white circle in Fig. S8b) and $E_{i0}$ is the energy on the parabolic curve at the $|\boldsymbol{q}|$ value corresponding to $E_i$. The addition of 50 meV in the denominator is just to get rid of negative value in the summation of $\chi^2$.

## Supplementary Figures and Tables

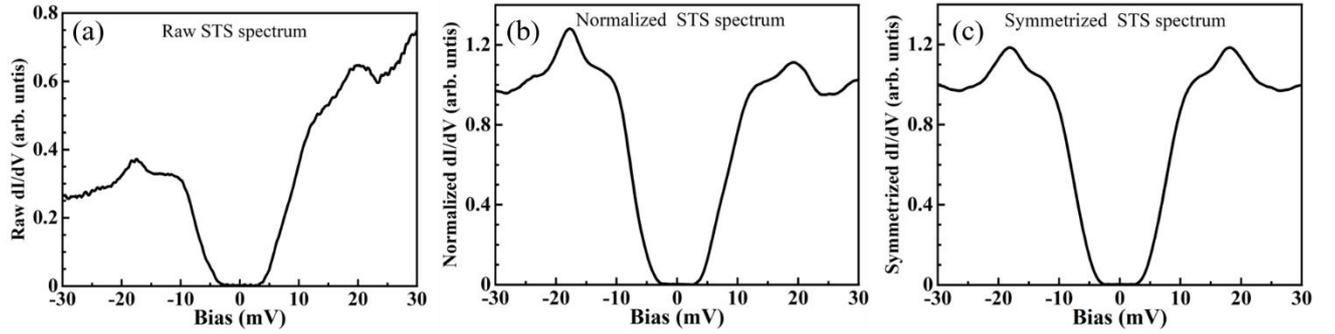

**Fig. S1** One typical example of STS data processing, showing (a) the raw STS data, (b) STS after normalization, (c) STS after normalization and symmetrization.

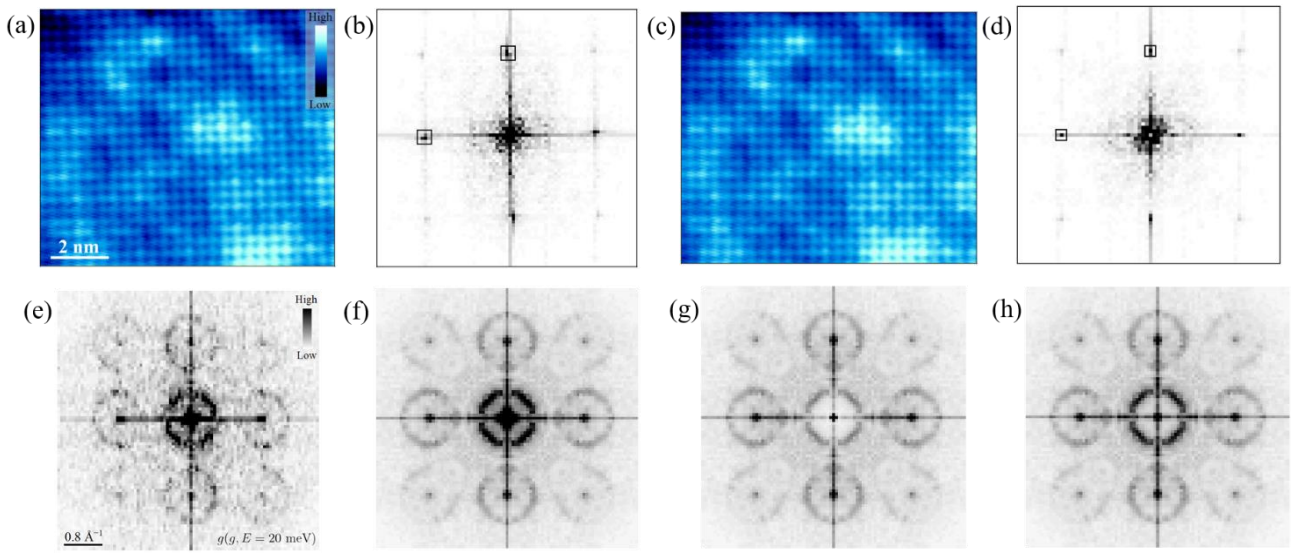

**Fig. S2 FT-QPI pattern measured in the uniform area of FeSe monolayer**. (a)-(b) Topographic images before the Lawler-Fujita correction and the corresponding FT pattern. (c)-(d) Topographic images after the Lawler-Fujita correction and the FT pattern. (e) $|g(q, E = 20 \text{ meV})|$ as the FT of the Lawler-Fujita corrected differential conductance mapping $g(r, E = 20 \text{ meV})$. (f) Mirror symmetrized and $C_4$ symmetrized $|g(q, E = 20 \text{ meV})|$. (g) $|g(q, E = 20 \text{ meV})|$ normalized by $|g(q, E = 0\ meV)|$. (h) $|g(q, E = 20 \text{ meV})|$ with a Gaussian core subtraction.

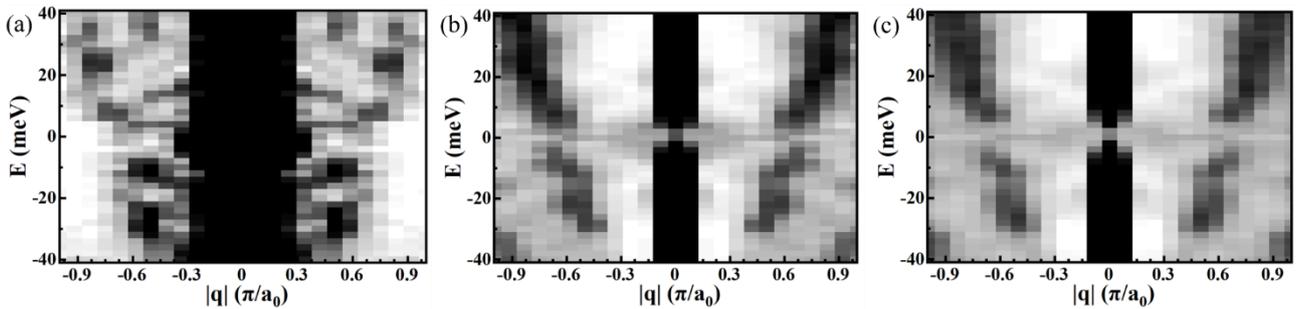

**Fig. S3 Band extraction from the uniform area of FeSe monolayer**. (a) Line cuts along $\theta = 22.5°$ of initial $|g(q, E)|$. (b) Line cuts along $\theta = 22.5°$ of symmetrized and normalized $|g(q, E)|$. (c) Azimuthally averaged line cuts of symmetrized and normalized $|g(q, E)|$.

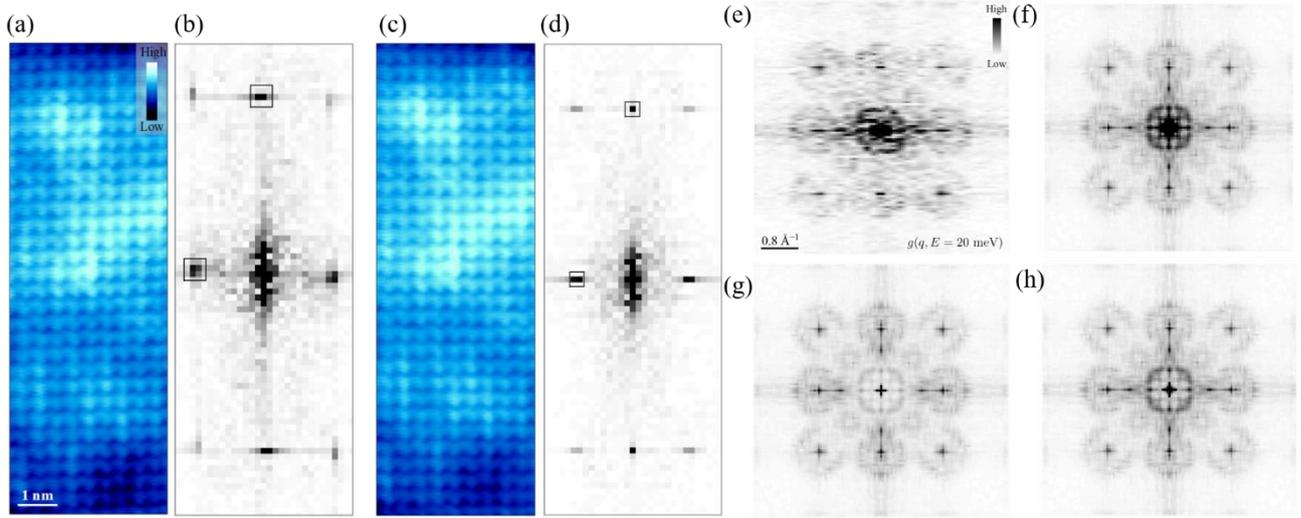

**Fig. S4 FT-QPI pattern measured in the FeSe peninsula.** The data process is the same as that in the uniform area of FeSe monolayer (Fig. S2).

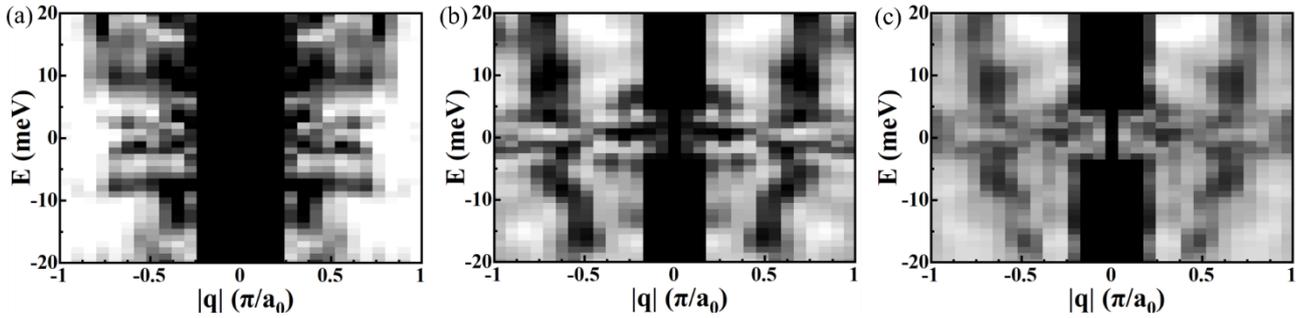

**Fig. S5 Band extraction measured in the FeSe peninsula.** The data process is the same as that shown in Fig. S3.

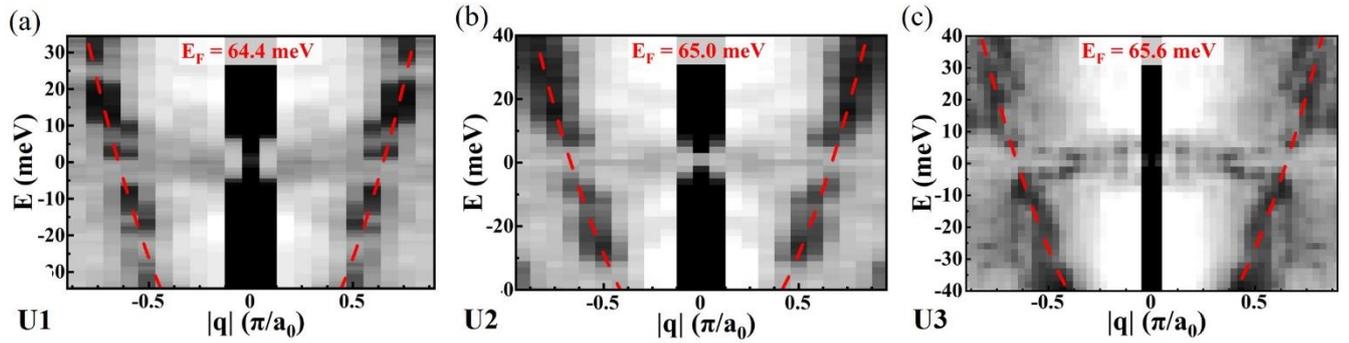

**Fig. S6** Band dispersions obtained from FT-QPI in uniform area (labeled as U1-U5) in the monolayer FeSe. The red dashed curve is the parabolic fitting to obtain $E_F$. U3 is shown as Figure 1(d) in the main text.

**Table S1** $E_F$ and $\Delta_{1,\,2}/E_F$ values in the uniform area of monolayer FeSe.

|  | U1 | U2 | U3 |
|---|---|---|---|
| $E_F$ (meV) | 64.4 | 65.0 | 65.6 |
| $\Delta_1/E_F$ | 0.16 | 0.16 | 0.15 |
| $\Delta_2/E_F$ | 0.28 | 0.28 | 0.28 |

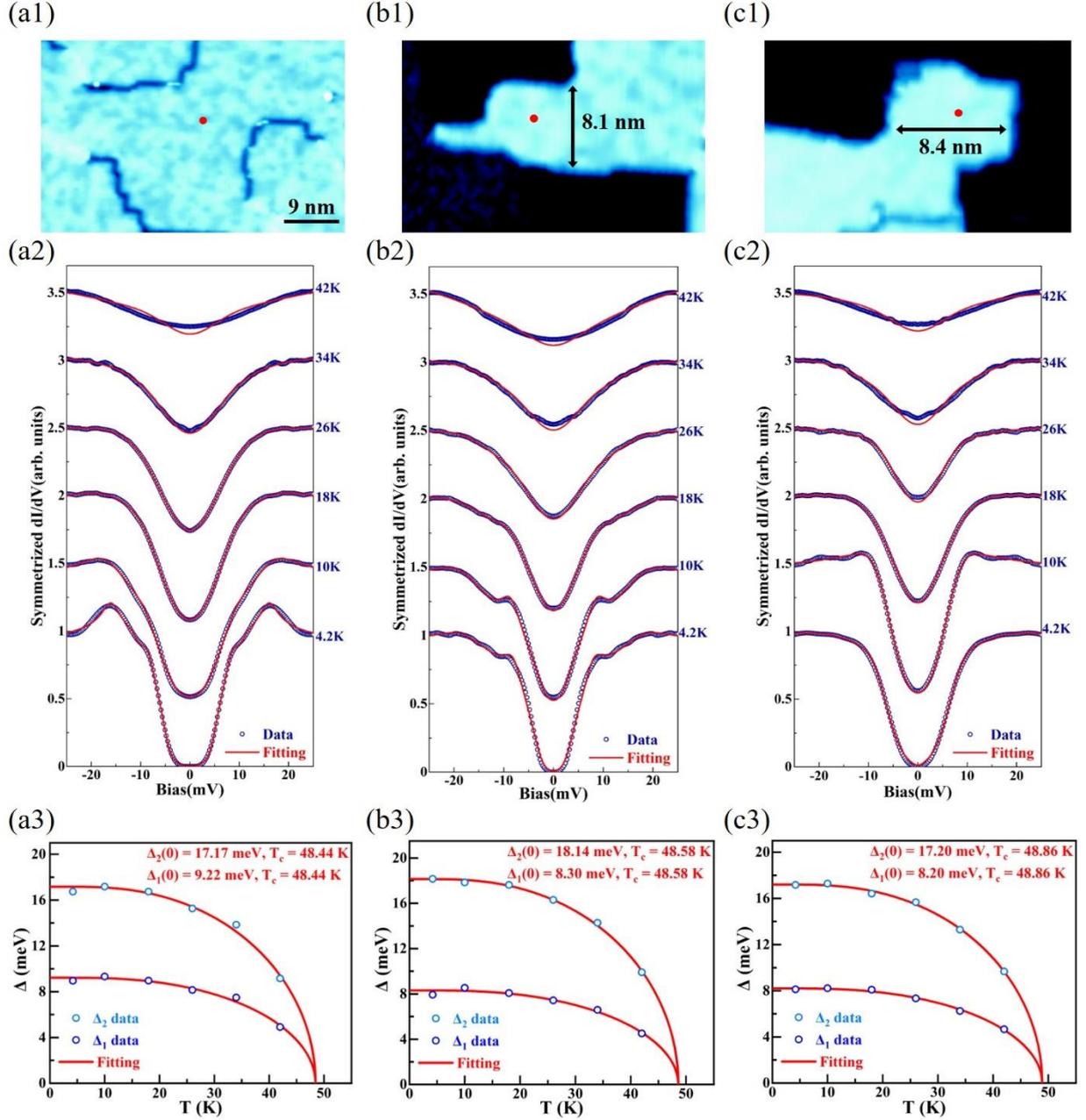

**Fig. S7** (a1) STM topographic image of a central uniform area in the monolayer FeSe, where the red dots denote the location of STS measurements. (a2) Temperature-dependent tunneling spectra (blue dots) with double-band anisotropic Dynes fittings (red curves). The spectra are vertically offset for clarity. (a3) BCS fittings (red curves) of $\Delta_{1,\,2}(T)$ values (blue dots) to obtained $T_c$. (b1-b3) Similar images and results to (a1-a3), but from a FeSe peninsula at the film boundary. (c1-c3) Images and results from another peninsula.

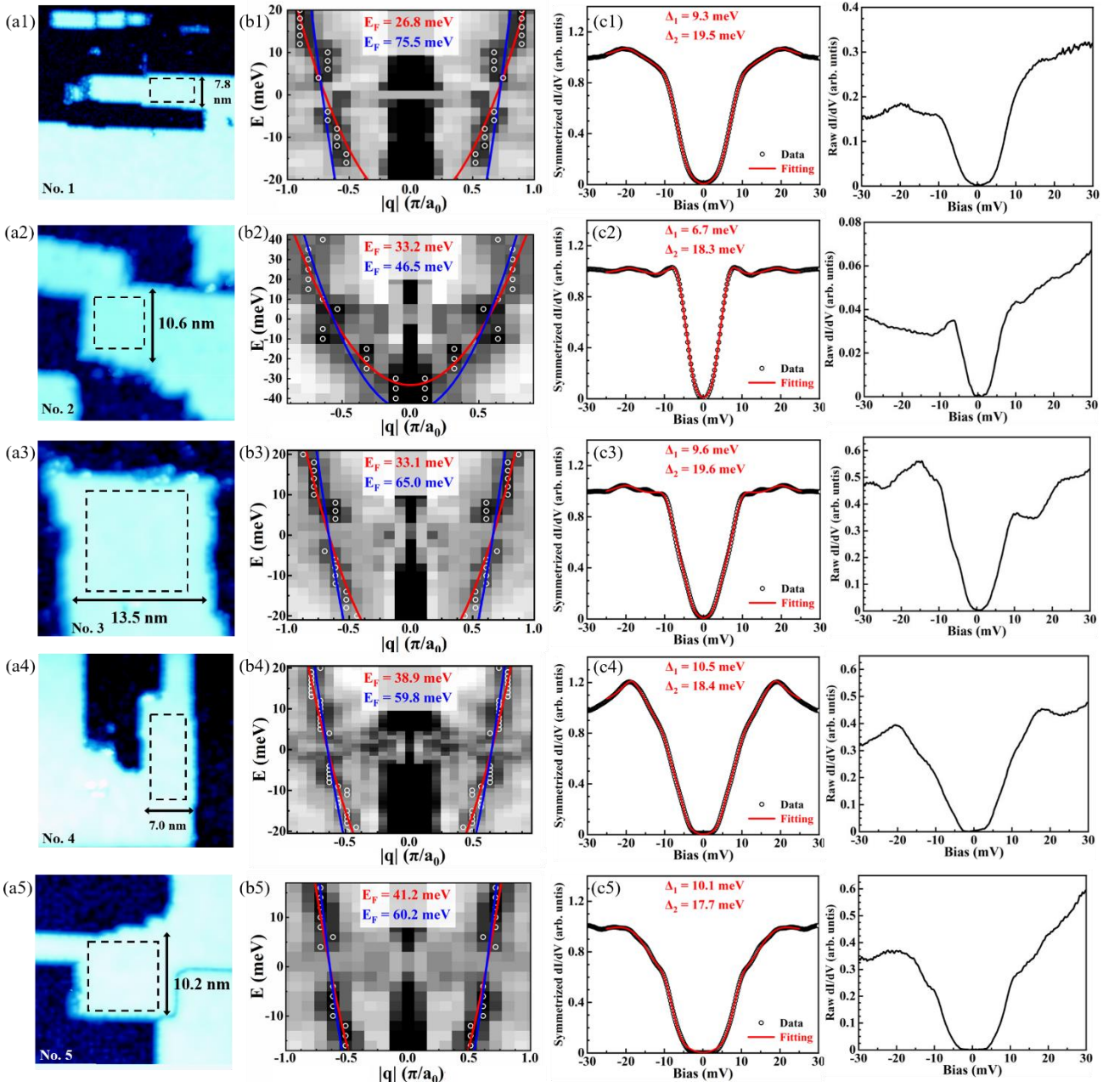

**Fig. S8** (a) STM topographies, (b) Band dispersions from QPI measurements, $E_F$ values are found by fittings with free $m^*$ parameter (red parabola) and fixed $m^* = 4m_e$ (blue parabola) (c) Tunneling spectra in peninsulas at the boundary of monolayer FeSe/STO, with the raw dI/dV data before normalization and symmetrization shown on the right. (a1) and (a4) correspond to Figure 2(a) and 2(d) in the main text.

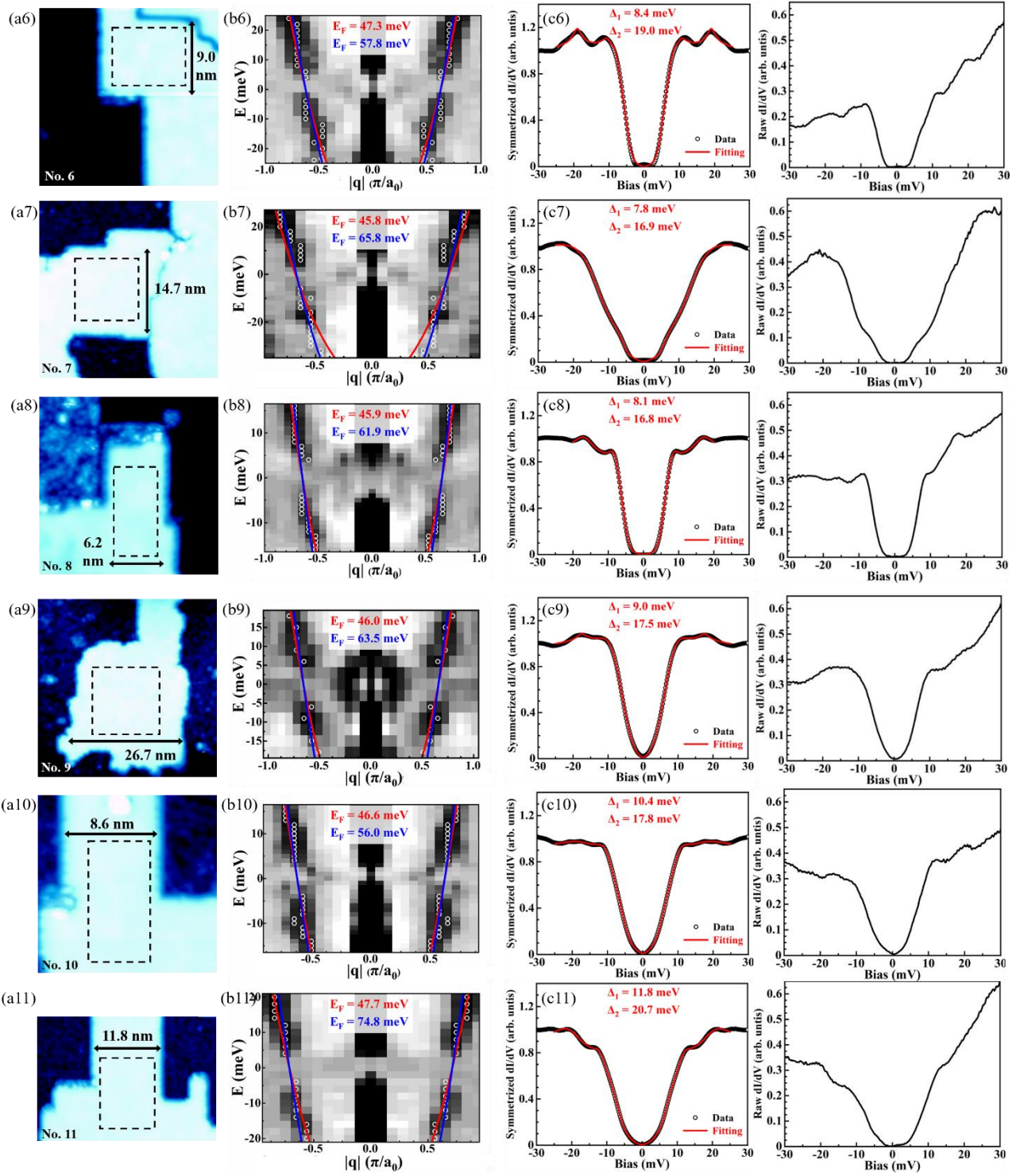

**Fig. S9** Results following Fig. S8. (a) STM topographies, (b) Band dispersions from QPI measurements and (c) tunneling spectra in other peninsulas at the boundary of monolayer FeSe/STO.

**Table S2** All results of $E_F$ and $\Delta_{1,2}/E_F$ values in FeSe peninsulas.

| No. | 1 | 2 | 3 | 4 | 5 | 6 | 7 | 8 | 9 | 10 | 11 |
|---|---|---|---|---|---|---|---|---|---|---|---|
| $E_F$ (meV) | 26.8 | 33.2 | 33.1 | 38.9 | 41.2 | 47.3 | 45.8 | 45.9 | 46.0 | 446.6 | 47.7 |
| $\Delta_1/E_F$ | 0.35 | 0.20 | 0.29 | 0.27 | 0.25 | 0.18 | 0.17 | 0.18 | 0.20 | 0.22 | 0.25 |
| $\Delta_2/E_F$ | 0.73 | 0.55 | 0.59 | 0.47 | 0.43 | 0.40 | 0.37 | 0.37 | 0.38 | 0.38 | 0.43 |

**Table S3** $\chi^2$ value for each fitting in Fig. S8, S9.

| No. | 1 | 2 | 3 | 4 | 5 | 6 | 7 | 8 | 9 | 10 | 11 |
|---|---|---|---|---|---|---|---|---|---|---|---|
| $\chi^2$ of fixed $m^*$ fittings | 135.2 | 125.2 | 70.9 | 61.0 | 13.5 | 29.7 | 65.0 | 21.2 | 8.9 | 15.3 | 27.5 |
| $\chi^2$ of free $m^*$ fittings | 8.9 | 37.0 | 13.4 | 11.2 | 5.3 | 19.4 | 56.1 | 12.2 | 4.1 | 13.2 | 4.3 |

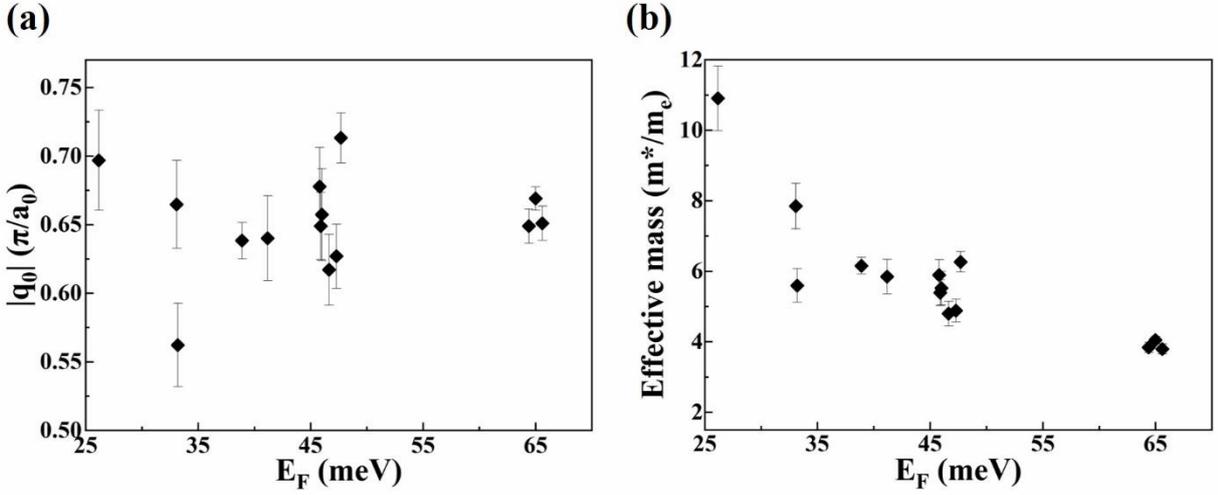

**Fig. S10** (a) $E_F$ in relation with the Fermi momentum of the band ($|\mathbf{k}_F| = |\mathbf{q}_0|/2$) (b) $E_F$ in relation with the effective mass $m^*$, in the unit of the electron mass $m_e$. Data come from the parabolic fittings for band dispersions in Fig. S6, S8 and S9, and error bars from the parameter uncertainty given by the fitting. Smaller $E_F$ values correspond to FeSe regions with more lateral confinements such as peninsulas, while larger $E_F$ values correspond to the uniform region.

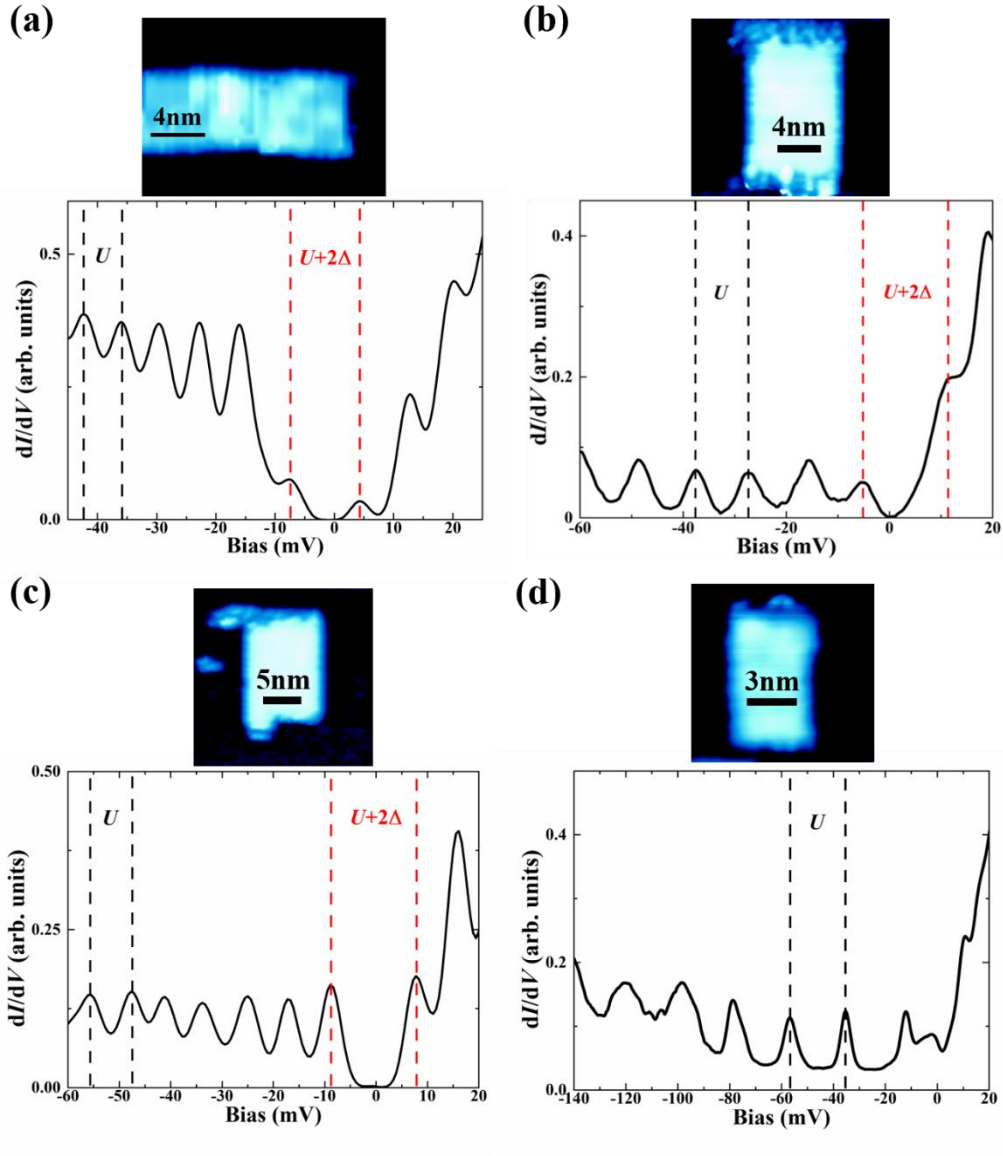

**Fig. S11** Typical topographic images and corresponding tunneling spectra for various monolayer FeSe islands with different volume ($V$). (a) $V$ = 54.1 nm$^2$, $U$ = 6.6 meV. (b) $V$ = 45.1 nm$^2$, $U$ = 11.1 meV. (c) $V$ = 64.4 nm$^2$, $U$ = 7.7 meV. (d) $V$ = 23.9 nm$^2$, $U$ = 21.0 meV. The superconducting gap is absent in this island with the volume smaller than the Anderson limit.

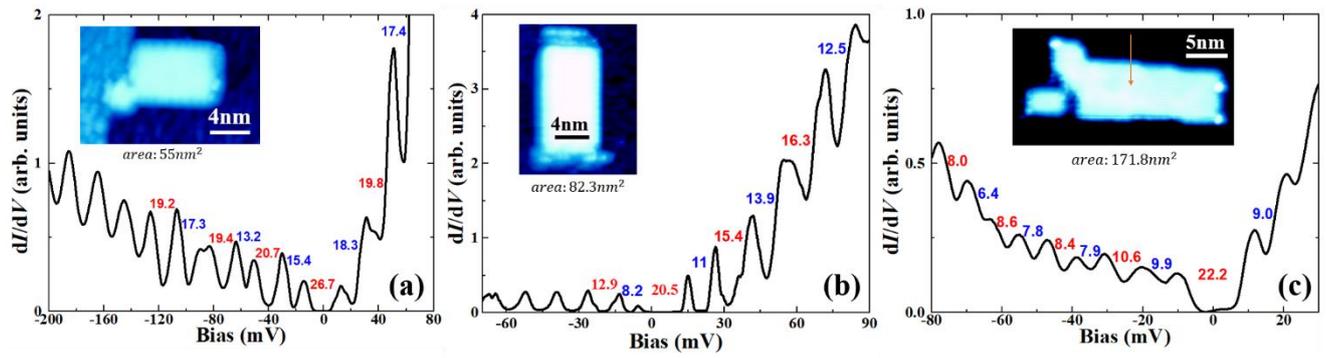

**Fig. S12** Topographic images and corresponding tunneling spectra for three monolayer FeSe islands with possible signatures of the even-odd effect. In the spectra, the number between every two conductance peaks indicates the distance between the peaks. Numbers in red (blue) are larger (smaller) than their neighboring blue (red) numbers, and these two colors alternate.